\documentclass[a4paper,11pt]{article}
\usepackage{pos}

\usepackage[switch]{lineno}
\usepackage{hyperref}

\usepackage{orcidlink}

\renewcommand{\logo}{\relax}

\title{LiquidO: Neutrino Detection and Imaging in Opaque Media}
\ShortTitle{LiquidO: Neutrino Detection and Imaging in Opaque Media}

\author[a]{Diana~Navas\mbox{-}Nicol\'as\,\orcidlink{0000-0002-2245-4404}\,}
\author[b]{Cloé~Girard\mbox{-}Carillo}
\author*[c]{Stefan~Schoppmann\,\orcidlink{0000-0002-7208-0578}\,}
\onbehalf{on behalf of the LiquidO Collaboration}

\affiliation[a]{CIEMAT, Centro de Investigaciones Energ\'{e}ticas, Medioambientales y Tecnol\'{o}gicas,\\28040~Madrid, Spain}
\affiliation[b]{Johannes~Gutenberg-Universität~Mainz, Institut für Physik, 55128~Mainz, Germany}
\affiliation[c]{Johannes~Gutenberg-Universität~Mainz, Detektorlabor, Exzellenzcluster~PRISMA${}^+$, 55128~Mainz, Germany}

\emailAdd{LiquidO-Contact-L@in2p3.fr}

\abstract{For several decades now, scintillator detectors have found a wide range of applications in particle physics, including neutrino detection, the search for dark matter and even medical imaging. These detectors so far have strongly relied on the transparency of the scintillating medium, through which light is typically propagated to surrounding photosensors. In this work, we present the results of a 10 litre prototype based on a novel detection approach where an opaque scintillator medium is used to confine light near its creation point that is then collected by a grid of wavelength-shifting fibres traversing the detector. The prototype is operated with different media, including the novel opaque scintillator NoWaSH whose scattering length varies with temperature. Our results progressively demonstrate the temperature-dependent stochastic confinement of the light, with 90\% (80\%) of the light being confined within a radius of 5\,cm (4\,cm) when the scattering length is on the order of a few millimetres. The results also demonstrate the pulse shape resolution of our setup capable of resolving Cherenkov and scintillation light. Altogether, the observations match the performance expected for this new type of detector, whose capabilities are expected to include the imaging of particle interactions down to MeV-energies.}

\FullConference{42nd International Conference on High Energy Physics (ICHEP2024)\\
18-24 July 2024\\
Prague, Czech Republic\\}

\begin{document}
\maketitle

\section{Introduction}\label{sec:intro}

For many decades now, optical detectors have played a central role in a wide range of physics disciplines spanning from astronomy and particle physics to nuclear physics and medical imaging~\cite{Shwartz:2017efz}. In these detectors, the energy and momentum of particles is converted into optical signals that are then detected by sensitive photosensors. In scintillator detectors, the energy loss of charged particles via ionisation results in the isotropic emission of fluorescence light over periods of time typically elapsing over hundreds of nanoseconds. In contrast, Cherenkov detectors exploit the fast light emitted by the Vavilov-Cherenkov effect in a cone centred around the direction of the charged particle, typically over tens of nanoseconds.

Essentially all implementations of these detectors to date have relied on the high transparency of their media to transport the light to photosensors, which are typically placed on the periphery. This proceedings presents the most recent experimental demonstration of a novel approach called LiquidO that exploits opacity. In this approach, which is described in detail in Ref.~\cite{LiquidO_2019,LiquidO_2024_LIME,LiquidO_2024}, the detector consists of a volume filled with opaque scintillator traversed by a dense array of fibres running in at least one dimension read out by photosensors. The opaque scintillator that is ideal for this type of detector has a long absorption length, but a short scattering length~\cite{Buck_2019}. The opacity of the medium causes scintillation and Cherenkov light to undergo a stochastic random walk, resulting in the creation of a so-called ``light ball'' around its creation point. This behaviour allows to preserve some of the topological information of particle interactions and endows the detector with powerful particle identification capabilities down to the MeV-scale. At these energies, electrons appear as single light balls created by their very short ionisation trail, while gamma-rays show up as chains of light balls produced from their Compton scatters. Positrons combine both pattern due to their two annihilation gammas~\cite{LiquidO_2019}.

In this proceedings, the prototype introduced in Ref.~\cite{LiquidO_2024}, that we refer to as Mini-LiquidO, is used. Electrons from a beam spectrometer are used to study the behaviour of the detector with point-like depositions of known energy. Section~\ref{subsec:setup} describes the experimental setup and the data collected. Section~\ref{subsec:confinement} shows how the data demonstrate stochastic light confinement around the beam spot. Finally, Section~\ref{subsec:timing} discusses the pulse shape capabilities. Additional details about the setup, the reconstruction, and the measurements are reported in Ref.~\cite{LiquidO_2024}.

\section{Setup and Data Acquisition}\label{subsec:setup}
\begin{figure}[t]
\centering
    \includegraphics[width=1.0\linewidth]{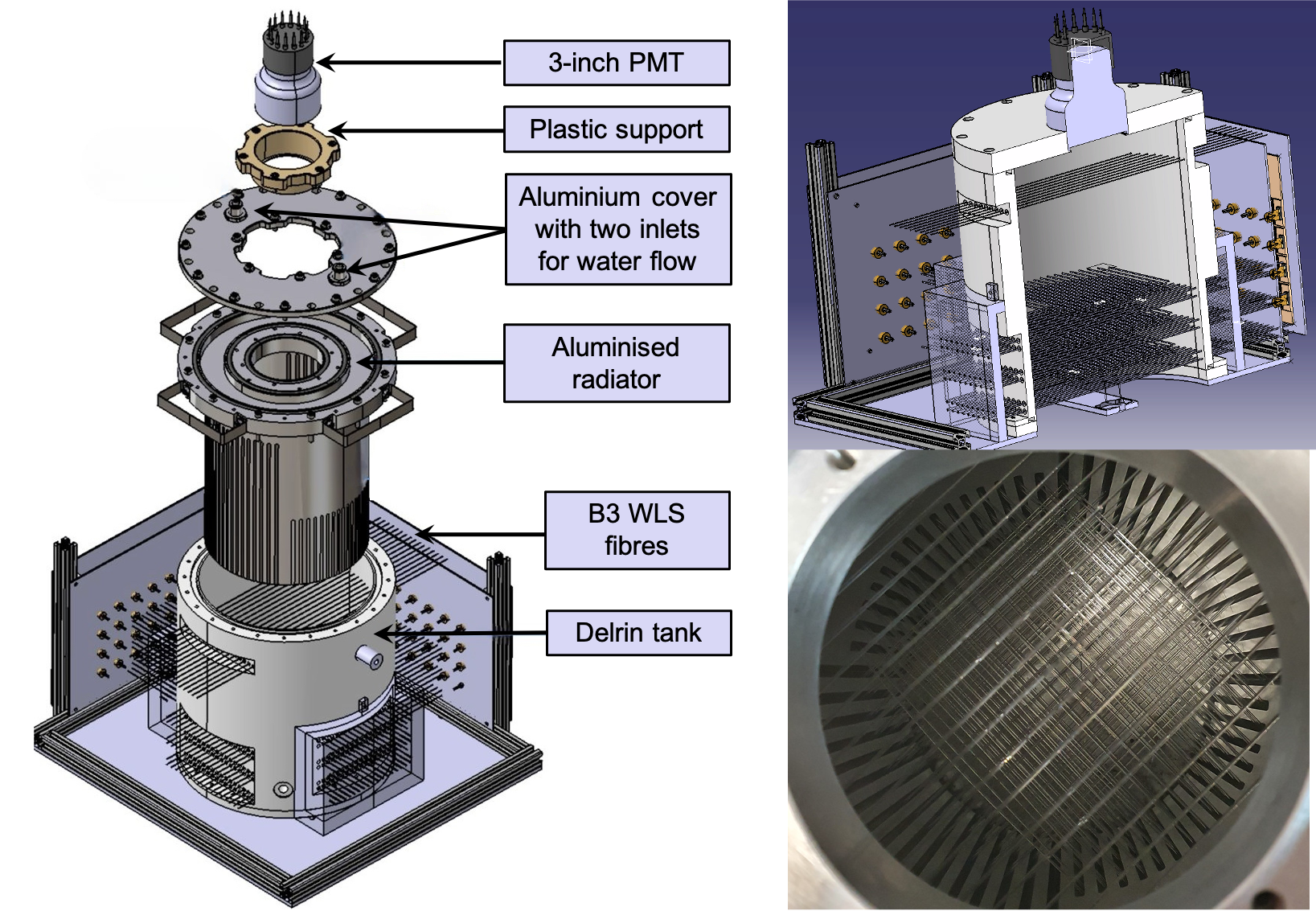}
    \caption{Mini-LiquidO experimental setup. Left: exploded-view drawing, top right: cutaway drawing, bottom right: photograph of the inner part taken through the opening for the PMT on top, while the PMT was removed~\cite{LiquidO_2024}. See \autoref{subsec:setup} for details.}
    \label{fig:setup}
\end{figure}

We perform our measurements in the Mini-LiquidO prototype shown in \autoref{fig:setup}, which is a cylindrical vessel of approximately 10 litres of volume.
We make use of two transparent detector media and one opaque medium.
The transparent media are water and linear alkylbenzene (LAB) scintillator with and without the addition of diphenyloxazole (PPO) as fluorophore.
The opaque medium is a scintillator consisting of LAB mixed with paraffin wax and PPO at a concentration of 3g/l, that is referred to as NoWaSH~\cite{Buck_2019}.
We use NoWaSH at 20\% wax loading by weight (NoWaSH-20), which can be set between 5\,\textcelsius, the temperature at which it is opaque, and 40\,\textcelsius, at which it is transparent.
Light readout is performed simultaneously by a 3-inch PMT (HZC photonics XP72B20) embedded in the plate at the top of the fiducial volume and 208 Kuraray B-3 wavelength-shifting fibres traversing the volume along two orthogonal directions parallel to the plane of the floor, of which 56 fibres are coupled to a Hamamatsu S13360--1350PE silicon photomultiplier (SiPM) on one end while the other end is left open. Signal readout for each channel is performed by custom front-end and digitisation hardware with a time resolution below 100\,ps.
A more detailed description of the Mini-LiquidO detector is available in Ref.~\cite{LiquidO_2024}.
Mini-LiquidO is positioned in the electron beam of a high energy resolution spectrometer situated at LP2i in Bordeaux, France.
The narrow-energetic beam is tuned to energies between 0.4 and 1.8\,MeV.
The beam impinges from the bottom, its direction coinciding with the axis of the cylindrical vessel and is externally triggered by tagging the single-electron with a thin plastic scintillator.

We collected different data sets with pure water, pure LAB and LAB loaded with 3 g/l PPO at 20\,\textcelsius~ as well as with NoWaSH-20 at varying temperatures between 5\,\textcelsius~and 40\,\textcelsius~.
For these runs, the electron beam energy was varied between 0.4 and 1.8\,MeV.
Since the energy of the electrons is low, they deposit all their energy in a small volume of a few cubic millimetres very close to their entry point at the bottom.
We therefore can assume a point-like source of light. 

The results reported throughout this proceedings are obtained by comparing our data sets in a relative fashion, using the transparent media (water, LAB and the high-temperature NoWaSH) as our references. 
This approach allows us to be independent of detector simulations and eliminates systematic effects due to different setups.

\section{Stochastic Light Confinement}\label{subsec:confinement} 
\begin{figure}[t]
    \includegraphics[width=0.75\linewidth]{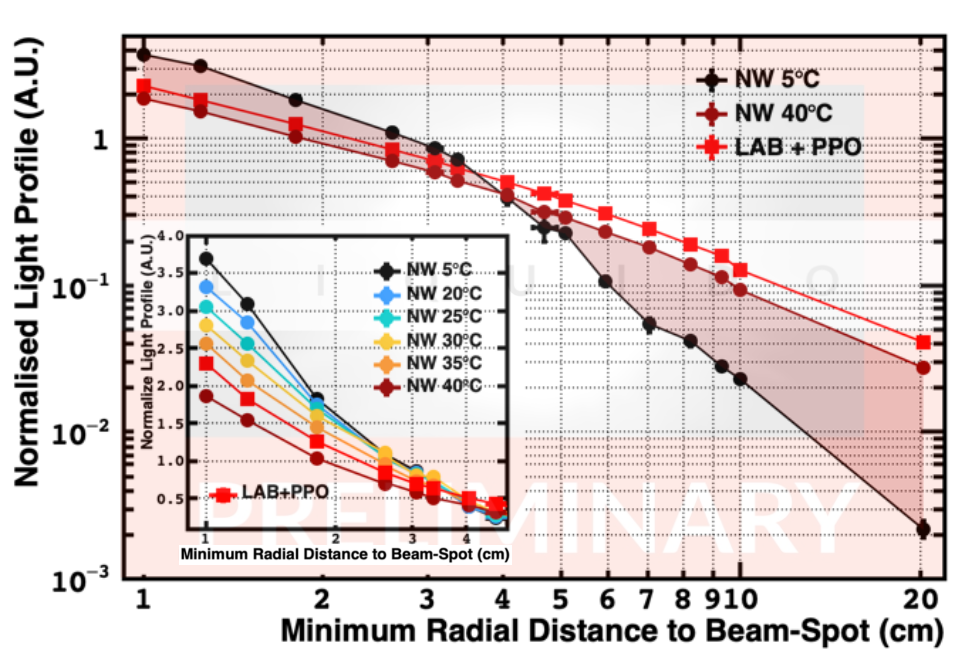}
    \centering
    \caption{Amount of light detected after noise subtraction and efficiency correction as a function of the minimal distance between each fibre and the beam spot. Fibres with similar minimal distance are clustered into single data points.
    NW stands for NoWaSH at 20\% wax-loading (NoWaSH-20), and is the only medium whose temperature is systematically varied. The inset shows the results for various intermediate NoWaSH-20 temperatures between 5\,\textcelsius~and 40\,\textcelsius, illustrating how the light confinement occurs gradually.}
    \label{fig:lyorbitals}
\end{figure}

A key goal of this prototype is to study if and how light is stochastically confined near its creation point as a result of the opacity of the medium.
A theoretical description of this confinement process is given in Ref.~\cite{LiquidO_2024}.
\autoref{fig:lyorbitals} shows the amount of light detected per cluster of fibres as a function of the minimal distance to the 1.8 MeV electron injection point, for different media and temperature configurations. Even though every fibre samples the light distribution across different distances from the beam spot, this minimal distance is obtained by measuring it in the direction perpendicular to the fibres. Fibres with a similar minimal distance are clustered into a single point by averaging their light yields and minimal distances. The amount of light detected per channel is obtained as the sum of the total number of photoelectrons (PE) observed after subtracting noise events and correcting for inefficiencies, driven primarily by channel-wise differences in the optical coupling between the fibres and the SiPMs as detailed in Ref.~\cite{LiquidO_2024}, and divided by the total number of events. A common normalisation factor is applied to all the curves.

It is worth noting that the transparent NoWaSH-20 at 40\,\textcelsius~curve is consistently lower than the LAB+PPO curve at ambient temperature.
This is to be expected due to the higher temperature and the non-scintillating material (paraffin wax) present in NoWaSH-20.

Comparing data from transparent and opaque configurations shows that approximately 65\% more light is detected in the fibres closer to the beam spot with the NoWaSH-20 at 5\,\textcelsius~while significantly less light is detected in those fibres farther than about 4\,cm.
We find 90\% (80\%) of the light being confined within a radius of 5\,cm (4\,cm).
The effect is reproducible for various beam energies and is independent of the details of the relative efficiency correction.

The results show the formation of the so-called light ball resulting from the stochastic confinement of the light around the beam spot.
The decrease of the scattering length of the NoWaSH-20 sample, which is achieved by decreasing its temperature, results in a significantly different light collection profile whereby increasingly more light is detected in the fibres closer to the beam spot compared to the farthest ones.
The increase in light collection by the closer fibres in the low temperature NoWaSH-20 compared to the LAB+PPO scenario (or the NoWaSH-20 at 40\,\textcelsius~scenario) confirms that the light is not being lost and rules out an absorption-only scenario.
These observations confirm the initial findings of Ref.~\cite{LiquidO_2019}. 

\section{Pulse Shape}\label{subsec:timing}
\begin{figure}[t]
    \includegraphics[width=0.75\linewidth]{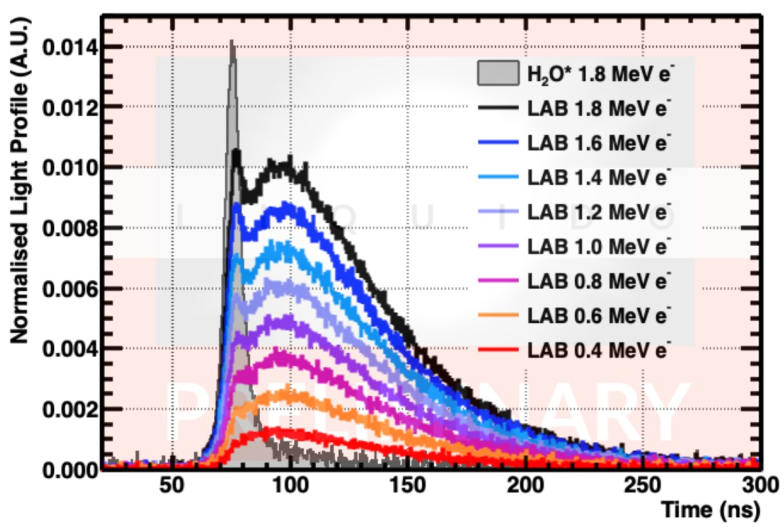}
    \centering
    \caption{Separation of light from Cherenkov and scintillation. An electron beam of varying kinetic energy between 0.4 and 1.8\,MeV is producing light in Mini-LiquidO filled with pure transparent LAB without fluorophores. In this material, the threshold for the production of Cherenkov light by electrons lies around 0.2\,MeV of kinetic energy. For energies above, the fraction of fast Cherenkov light is increasing compared to the bulk of slow scintillation light, as a function of the electron energy, as expected. The resolution of the read-out electronics of the detector allows a clear separation between both light pulses. In comparison, the pure Cherenkov light pulse when operating Mini-LiquidO with non-scintillating water is shown for an electron kinetic energy of 1.8\,MeV. We correct its height for the differing refractive index~\cite{LiquidO_2024}. The time structure of the pulse is clearly matching the Cherenkov component of the LAB.}
    \label{fig:timing}
\end{figure}
An additional handle to identifying particles comes from exploiting the particle-dependent fraction of Cherenkov and scintillation light produced.
Hybrid detector media (e.g.~water-based or slow scintillators) are in fact being developed that would allow to exploit this capability~\cite{Schoppmann_2023}.  
In both cases, separating the two types of light becomes possible e.g.~by exploiting their distinct production time profiles.  

To illustrate the separation power that is achievable with a LiquidO-type detector, we fill the Mini-LiquidO prototype with pure water as well as pure LAB without fluorophores, which has an intrinsically slow time constant.
As seen in \autoref{fig:timing}, we show that light separation is possible thanks to the excellent timing resolution of SiPMs and the fast electronics coupled to them.
While the use of pure LAB results in a low light yield, the addition of low concentrations of fluorophores, or intrinsically slow fluorophores or solvents, will increase the light yield at similar separation power~\cite{Schoppmann_2023}.
These results are transferable to opaque scintillators~\cite{NuDoubt}. 

\section{Conclusion}\label{sec:conclusion}
This proceedings presents the results obtained with a 10 liter prototype of the LiquidO approach~\cite{LiquidO_2024}.
This is the largest prototype that has been assembled to date for this technology.
Using a high-resolution electron beam spectrometer that produces point-like energy depositions at the bottom of the detector, as well as NoWaSH scintillator whose scattering length decreases with temperature, the stochastic confinement of the light near the beam spot was conclusively demonstrated by comparing the light detection profile across the detector at various temperatures.
The combination of a fast SiPM-based light readout system with a slow scintillator also allowed to separate the Cherenkov and scintillation light produced from the electron energy depositions.
A larger LiquidO prototype is already under construction that will be used to explore the particle identification capabilities of this technology in detail.

\end{document}